\documentclass[11pt]{article}
\pdfoutput=1
\usepackage{jcappub}
\usepackage{slashed}
\usepackage{stackengine}
\usepackage{amsmath}
\usepackage[frak=euler,scr=boondox,bb= pazo]{mathalfa}

\def\be{\begin{equation}}
\def\ee{\end{equation}}
\def\bea{\begin{eqnarray}}
\def\eea{\end{eqnarray}}

\begin{document}

\title{Emergent Quantumness in Neural Networks}

\author[a]{Mikhail I. Katsnelson}

\emailAdd{m.katsnelson@science.ru.nl}

\author[b]{and Vitaly Vanchurin}

\emailAdd{vvanchur@d.umn.edu}

\date{\today}

\affiliation[a]{Institute for Molecules and Materials, Radboud University, Heyendaalseweg 135, NL-6525 AJ Nijmegen, The Netherlands}

\affiliation[b]{Department of Physics, University of Minnesota, Duluth, Minnesota, 55812 \\
Duluth Institute for Advanced Study, Duluth, Minnesota, 55804}

\abstract{
It was recently shown that the Madelung equations, that is, a hydrodynamic form of the Schr\"odinger equation, can be derived from a canonical ensemble of neural networks where the quantum phase was identified with the free energy of hidden variables. We consider instead a grand canonical ensemble of neural networks, by allowing an exchange of neurons with an auxiliary subsystem, to show that the free energy must also be multivalued. By imposing the multivaluedness condition on the free energy we derive the Schr\"odinger equation with ``Planck's constant'' determined by the chemical potential of hidden variables. This shows that quantum mechanics provides a correct statistical description of the dynamics of the grand canonical ensemble of neural networks at the learning equilibrium. We also discuss implications of the results for machine learning, fundamental physics and, in a more speculative way, evolutionary biology. }

\maketitle

\section{Introduction}

Despite the obvious success of quantum mechanics in description of our physical world, its conceptual status is still a subject of hot debates, see Refs. \cite{Home,Weinberg,Khrennikov,Rauch,Landsman}, to name just a few contemporary books; more references can be found in the recent papers \cite{separation,neural_network}. As a result, many so-called ``no-go theorems'' were constructed (e.g. Bell's inequalities \cite{Bell}) to rule out the existence of a hidden classical world beyond quantum mechanics \cite{Bohm}. Recently the idea of ``emergent quantumness'' was reincarnated in the programs like ``the world as a matrix'' \cite{Adler}, ``the world as a cellular automaton'' \cite{tHooft1,tHooft2} and ``the world as a neural network'' \cite{neural_network, machine_learning}. An alternative approach was suggested in the ``logical inference'' program \cite{logical_inference1,logical_inference2,logical_inference3}, where quantum mechanics is considered as a purely phenomenological way to describe the results of repeated experiments assuming that (1) we cannot control all the details relevant for these experiments, (2) our description should be as robust as possible, and (3) it should follow some ``axioms of rational thinking''. From this point of view, the question on the existence of the hidden world beyond quantum is claimed to be irrelevant: whatever this world is, we are forced, by the properties of our mind, to describe the reality via something similar to quantum theory.

At a phenomenological level,  the  ``neural network''  \cite{neural_network} and the ``logical inference'' \cite{logical_inference1,logical_inference2,logical_inference3} approaches are not contradictory, and may in some sense be dual to each other. Indeed, if our mind could be modeled as a neural network, then it is not too unreasonable to expect that the principles of work of the neural network \cite{machine_learning} might be used to derive the postulated axioms of the logical inference \cite{Cox}. This possibility is supported by the fact that in both approaches one is able to derive Schr\"odinger equation \cite{logical_inference1,neural_network} by combination of some entropic variational principle aimed to provide the most robust and efficient description of the external world with some form of the Hamilton-Jacobi equations (see also Refs. \cite{Caticha, entropic} for other derivations based on entropic principles.) There is, however, an important flaw in these constructions, explicitly mentioned in Ref.\cite{logical_inference1}. What is actually derived, in both cases, is not the Schr\"odinger, but  the Madelung \cite{Madelung} hydrodynamic equation which is known to be different from the Schr\"odinger equation \cite{Wallstrom}. The key difference is in the global topology. In the Madelung form of the Schr\"odinger equation, we introduce the ``fluid density'' which is related, in quantum language, to the modulus of the wave function, and the ``fluid velocity'' which is related to a gradient of the phase of the wave function. However, in the Schr\"odinger equation the phase is defined modulo 2$\pi$ (we glue the plane to a cylinder) and in the Madelung equation this condition is lost. Without it, the Madelung hydrodynamics describes only a very special kind of hydrodynamic flows, that is, curlless (without vortices) whereas the crucial point of quantum physics, the quantization (in particular, discreteness of atomic energy levels) is associated with discreteness of circulation, like in superconductors and superfluids \cite{Feynman,Abrikosov}.\footnote{MIK thanks Grigory Volovik for emphasizing this connection at our old discussions of the logical inference approach.} In some inexplicit way, this change of topology simplifies the description allowing to pass from the {\it nonlinear} Madelung equations to the {\it linear} Schr\"odinger equation adding extremely powerful machinery of vectors and operators in a Hilbert space. Phenomenologically, it can be justified by introducing a new principle of ``separation of conditions'' \cite{separation}, logically independent from the logical inference approach. However, due to extreme importance of this point one needs to have a more detailed understanding of its origin. In this paper we will show that the neural network approach gives a natural way to understand this transition and thus allowing to understand deeper the origin and meaning of one of the most fundamental physical constants, namely, the Planck constant.

The other point of a great interest is a very controversial, but tempting, idea of ``emergent quantumness'', that is, some quantum-like behavior of systems which are difficult to believe to be quantum per se \cite{Khrennikov1}. Some authors use quantumness just as a metaphor to describe cultural phenomena \cite{Wings} or genotype-phenotype duality in biological evolution \cite{Koonin}, while others suggest the relevance of the true quantum phenomena in human brains  \cite{Penrose,Fisher}. On a more practical level, this line of thinking may be related to a much more pragmatic and solid concept of ``quantum annealing'' \cite{QA1,QA2,QA3,QA4,QA5,QA6}. For example, if some optimization problem can be stated as the problem of finding a ground state of a complicated {\it classical} system, then it is often convenient to add to the system some ``quantumness'' because the quantum tunneling would prevent a self-trapping of the optimization process in one of metastable states. We will show here that the neural network approach gives a natural explanation of the emergent quantumness and provides a solid formal background for speculations on quantum behavior of non-quantum (e.g. macroscopic and even biological) systems.

The paper is organized as follows.  In Sec. \ref{sec:entropy_production} we apply the principle of stationary entropy production to derive a functional which governs the emergent dynamics of neural networks. In Sec. \ref{sec:schrodinger} we argue that the dynamics can be described by the Schrodinger equation if and only if the free energy of the hidden variables is a multivalued function. In Sec. \ref{sec:grand_canonical} we construct a grand canonical ensemble of neural networks and show that the corresponding free energy is multivalued. In Sec. \ref{sec:superposition} we discuss in more detail some fundamental issues such as the difference between pure and mixed states, role of measurements and relations to path-integral formulation of quantum mechanics \cite{feynman_hibbs,schulman,kleinert}. In Sec. \ref{sec:discussion} we discuss implications of the main results for machine learning, physics and biology.

\section{Stationary entropy production}\label{sec:entropy_production}

Consider a learning system described by a coupled dynamics of {\it trainable} variables, ${\bf q}$,  and non-trainable or {\it hidden} variables, ${\bf x}$. In ``epistomological'' kind of approaches \cite{logical_inference1,logical_inference2,logical_inference3,separation} one can identify the trainable variables with characteristics of a human mind whereas the hidden variables represent an external world, but this identification is not needed for our formal consideration which we will try to keep as general as possible. In context of artificial neural networks the trainable variables determine the weight matrix and bias vector, and the hidden variables represent the state vector of neurons \cite{machine_learning}. It is assumed that on the shortest time-scales the dynamics of the trainable variables undergoes diffusion
\bea
\frac{\partial p(t, {\bf q})}{\partial t}&=&  \sum_k \frac{\partial}{\partial q_{k}}  \left (D  \frac{\partial p(t, {\bf q})}{\partial q_{k}} - \frac{d {q}_k}{dt} p(t, {\bf q}) \right)\notag\\
&=&  \sum_k \frac{\partial}{\partial q_{k}}  \left (D  \frac{\partial p(t, {\bf q})}{\partial q_{k}} - \gamma \frac{\partial F(t, {\bf q})}{\partial q_k}  p(t, {\bf q}) \right)
\label{eq:dP}
\eea
and the dynamics of hidden variables is only described through its free energy 
\bea
\frac{d}{dt} F(t, {\bf q}) &=&  \frac{\partial F(t, {\bf q})}{\partial t}  + \sum_k \frac{d q_k}{d t} \frac{\partial F(t, {\bf q})}{\partial q_k}\notag\\
 &=&  \frac{\partial F(t, {\bf q})}{\partial t}  +  \gamma \sum_k \left ( \frac{\partial F(t, {\bf q})}{\partial q_k} \right )^2     \label{eq:dF}
\eea
where the trainable variables experience a classical drift in the direction of the gradient of the free energy,
\be
 \frac{d q_{k}}{d t}  =  \gamma \frac{\partial F(t, {\bf q})}{\partial q_k}  \label{eq:dq}.
\ee
(See Ref. \cite{machine_learning, neural_network} for details.) We shall assume that the drift $\gamma$  and diffusion $D$ coefficients are constants (independent of ${\bf q}$), but their numerical values depend on a learning algorithm. For example, if a neural network is trained using the stochastic gradient descent, then  $\gamma$ and $D$ depend on the learning rate and the mini-batch size \cite{SGD}. Note that the system under consideration is supposed to be, initially, purely classical and subjected by stochastic and, moreover, dissipative dynamics described by equations (\Ref{eq:dP})-(\Ref{eq:dq}). One can think in particular on conventional neural network algorithms realized at normal classical computers, nothing specifically quantum is assumed yet.

 To describe the dynamics on longer time-scales we employ the principle of stationary entropy production: \\
\\
{\bf Principle of Stationary Entropy Production}: {\it The path taken by a system is the one for which the entropy production is stationary.}\\
\\
The principle was first introduced in Ref.\cite{entropic} as a generalization of both, the maximum entropy principle \cite{Jaynes, Jaynes2} and the minimum entropy production principle \cite{Prigogine,Klein}. In context of the neural networks the entropy production must be maximized in an optimal neural architecture  \cite{machine_learning, neural_network}.  The rationale behind it is simple: if less information is used for optimizing the network for past data, then more entropy is available for optimizing the network for future data and, therefore, a larger space of solutions can be explored. With this respect the principle can be thought of as a formalization of the Occam's razor principle.\footnote{We thank Nikolay Mikhailovsky for pointing out the connection to Occam's razor principle.}

 The Shannon entropy of the trainable variables is given by
 \be
{S}_q(t) = - \int d^K q \;\;{p}(t, {\bf q}) \log \left ( {p}(t, {\bf q}) \right ) \label{eq:entropy2}
\ee
 and the total entropy production can be calculated from \eqref{eq:dP},
\bea
\frac{d S_q(t) }{d t} & = &  - \int d^K q \;{p}  \frac{\partial  \log ({p}) }{\partial t }    - \int d^K q \; \log ({p})  \frac{\partial p}{\partial t }  \notag \\
& =&  - \frac{d}{dt} \int d^K q\; p    - \int d^K q \; \log ({p})  \frac{\partial p }{\partial t }  \notag \\
& =&    - \int d^K q \; \log ({p})\sum_k   \frac{\partial}{\partial q_{k}}  \left (D  \frac{\partial p}{\partial q_{k}} - \gamma \frac{\partial F}{\partial q_k}  p \right)  \notag \\
& =&    D \int d^K q \; \sum_k \frac{1}{p} \left (\frac{\partial p}{\partial q_{k}}  \right )^2  - \gamma \int d^K q \; \sum_k   \frac{\partial p}{\partial q_{k}} \frac{\partial F}{\partial q_k}. \label{eq:production0}
\eea
The two terms represent the entropy production due to stochastic and learning dynamics of the trainable variables. At a learning equilibrium (i.e. $\nabla F \sim 0$) the entropy production of the trainable variables due to learning is subdominant and the total entropy production of trainable variables \eqref{eq:production0} can be approximated as
\bea
\frac{d S_q(t) }{d t}  & \approx &    \int d^K q\; \sqrt{p}  \left (- 4 {D} \sum_k \frac{\partial^2}{\partial q_k^2}   \right)  \sqrt{p}. \label{eq:production}
\eea
Note that this term is nothing but Fisher information which determines metrics in the information space \cite{Amari} and plays an important role in the logical inference approach to quantum mechanics \cite{logical_inference1,logical_inference2,logical_inference3} and in the information theory approach to emergent gravity \cite{emergent_gravity}. Further away from the equilibrium the entropy production due to learning cannot be ignored and the dynamics is better described by a classical Hamiltonian mechanics with the free energy, $F$, identified with the Hamilton's principle function (see \cite{neural_network} for details on both classical and quantum limits).

The problem of optimization of the entropy production \eqref{eq:production} subject to a constraint \eqref{eq:dF} can be solved using the method of Lagrange multipliers by defining a functional \cite{neural_network}
\bea
{{\cal S}}[p, F, \lambda] &=&   \int_0^T dt \frac{d S_q }{d t} + \lambda \int_0^T dt d^K q  \;p \left (\frac{\partial F}{\partial t}  + \gamma \sum_k \left ( \frac{\partial F}{\partial q_k} \right )^2 +   \frac{V}{\epsilon}  \right ),\notag\\
&=&   \int_0^T dt\, d^K q  \,\sqrt{p}  \left ( -4 {D} \sum_k\frac{\partial^2}{\partial q_k^2}  +\lambda \frac{\partial F}{\partial t}  + \lambda \gamma \sum_k \left ( \frac{\partial F}{\partial q_k} \right )^2 + \lambda  \frac{V}{\epsilon} \right)  \sqrt{p},\label{eq:action}
\eea
where the total time-averaged free energy production pre unit time step $\epsilon$ is
\be
 V({\bf q}) \equiv -  \left \langle {\epsilon}  \frac{d}{dt} F(t, {\bf q}) \right \rangle_t. \label{eq:total_free}
 \ee
Again, the time step $\epsilon$ here is just a parameter of our neural network related to the rate of computation at a given realization of the neural network algorithm and is not related to any fundamental physical constants such as Planck time, etc. We postpone the discussion of the fundamental physics till the last section, meanwhile it is better to keep in mind just some more or less standard computations on more or less standard computers. 

Note that the logical inference approach \cite{logical_inference1} leads to a technically very similar formulation, the principle of robustness of description shown to be equivalent to the minimum of Fisher information, but the second requirement, correctness of the Hamilton-Jacobi equations at the average, was postulated as a property of our world, in spirit of Bohr's correspondence principle. The neural network approach \cite{neural_network} provides us an explicit model with the phenomenologically desired properties. Of course, strictly speaking, one cannot exclude existence of other models which lead to more or less the same phenomenology, but this model seems to be, in some sense, the most natural.

It is convenient to rewrite the functional \eqref{eq:action} as
\bea
{{\cal S}}[p, F, \hbar] =  \frac{\lambda}{\epsilon} \int_0^T dt\, d^K q  \,\sqrt{p}  \left ( -\frac{\hbar^2}{2 m} \sum_k\frac{\partial^2}{\partial q_k^2}  +   \frac{ \partial (\epsilon F) }{\partial t}  + \frac{1}{2m} \sum_k \left ( \frac{\partial (\epsilon F)}{\partial q_k} \right )^2 +  {V} \right)  \sqrt{p}\label{eq:action2}
\eea
where
\be
m \equiv \frac{\epsilon}{2 \gamma},
\ee
and
\be
\hbar \equiv \epsilon \sqrt{\frac{4D}{\gamma \lambda}} \label{eq:hbar}.
\ee
The constant (\ref{eq:hbar}) will play the role of the Planck constant in the further consideration but currently it is just a combination of some parameters characterizing our neural network, and it is not assumed to be either microscopic or fundamental. Generally speaking, different neural networks can have different ``Planck constants'', and one can, in principle, even assume a variable ``constant'' during the computation. 

The main difference between \eqref{eq:action} and \eqref{eq:action2} is that instead of solving the equations for $p$, $F$ and $\lambda$, we are now solving them for $p$, $F$ and $\hbar$.  The optimal solutions are obtained by setting all possible variations of \eqref{eq:action2} to zero
\bea
\frac{\partial}{\partial \hbar}  {{\cal S}}[p, F,\hbar] &=&  - \int_0^T dt\, d^K q \frac{\hbar}{m} \sqrt{p}  \sum_k\frac{\partial^2}{\partial q_k^2} \sqrt{p}  = 0 \label{eq:ent}\\
\frac{\delta}{\delta F}  {{\cal S}}[p, F,\hbar]&=& - \frac{\partial }{\partial t} p - \frac{1}{m} \sum_k \frac{\partial }{\partial q_k}  \left ( \frac{\partial  (\epsilon F)}{\partial q_k} p  \right )   =0 \label{eq:var_F}\\
\frac{\delta }{\delta p}  {{\cal S}}[p, F,\hbar] &=&-\frac{\hbar^2}{2 m} \frac{1}{\sqrt{p}}  \sum_k \frac{ \partial^2 \sqrt{p}}{\partial q_{k}^2}  + \frac{\partial  (\epsilon F)}{\partial t} + \frac{1}{2m}  \sum_k \left ( \frac{\partial  (\epsilon F)}{\partial q_k} \right )^2  + V =0.  \label{eq:var_p}
\eea

\section{Schr\"odinger dynamics}\label{sec:schrodinger}

In the previous section we derived the functional \eqref{eq:action2} which describes the total entropy production of the trainable variables ${\bf q}$ subject to a constraint imposed on the dynamics of free energy of the hidden variables ${\bf x}$. The stationary solutions for probability density, $p(t, {\bf q})$, free energy, $F(t, {\bf q})$ and ``Planck constant'', $\hbar$, are given by equation \eqref{eq:ent}, which represents conservation of entropy
\be
\left \langle \frac{d}{dt} S_q \right \rangle_t = 0, \label{eq:neural_SQ}
\ee
and  by equations \eqref{eq:var_F} and \eqref{eq:var_p}, which are the Madelung hydrodynamic equations \cite{Madelung}
\bea
\frac{\partial}{\partial t} p &=& - \sum_k \frac{\partial}{\partial q_k} \left ( u_k p \right )  \label{eq:neural_FP}\\
\frac{\partial}{\partial t} u_j &=& -   \sum_k u_k \frac{\partial}{\partial q_k} u_j  - \frac{1}{m} \frac{\partial}{\partial q_j} \left ( V  - \frac{\hbar^2}{2 m}  \sum_k \frac{ \partial^2 \sqrt{p}}{\partial q_{k}^2} \right )\label{eq:neural_NS}
\eea
with velocity of the fluid
\be
u_k \equiv \frac{1}{m} \frac{\partial}{\partial q_k} (\epsilon F). \label{eq:u}
\ee

It is well known that  the Madelung equations can be derived from the Schr\"odinger equation
\be
-  i \hbar \frac{\partial }{\partial t } \Psi  =  \left ( \frac{\hbar^2}{2 m} \; \sum_k \frac{ \partial^2 }{\partial q_{k}^2} -  V \right ) \Psi \label{eq:Schrodinger}
\ee
where the wave function is defined as
\be
\Psi \equiv \sqrt{p} \exp \left ( \frac{i F \epsilon}{\hbar} \right ). \label{eq:wavefunction}
\ee
This implies that the solutions of \eqref{eq:neural_FP} and \eqref{eq:neural_NS}  are also the solutions of the Schr\"odinger equation, \eqref{eq:Schrodinger}, and so it is expected that the time-averaged change in entropy is zero \eqref{eq:ent}, but the opposite is not true. The core of the problem is that the quantum phase in \eqref{eq:wavefunction} must be a multivalued function, but the free energy $F$ of hidden variables may or may not be single-valued. Therefore, in order to establish an equivalence between quantum mechanics and neural networks, we must consider statistical ensembles for which the free energy of hidden variables would be multivalued,
\be
F  \cong F + \mu n \;\;\;\;\;\;\; \;\;\;\;\;\;\;\forall n \in \mathbb{Z}.\label{eq:first_cond}
\ee
This can be accomplished by constructing a statistical ensemble of neural networks for which the discrete shift, $\mu n$, is not observable. In the following section we shall consider one such ensemble, the grand canonical ensemble with chemical potential, $\mu$, for which the exact number of neurons is unobservable and, therefore, the free energy is multivalued \eqref{eq:first_cond}. Note that the proportionality of thermodynamic potentials to the number of particles in a thermodynamic limit (a very large number of degrees of freedom), which is crucially important for our whole construction, can be proven mathematically rigorously for a broad class of continuous and lattice models of statistical mechanics \cite{Ruelle}.

If we assume for a moment that the multivaluedness condition \eqref{eq:first_cond} is  satisfied, then \eqref{eq:action2} can be rewritten as
\bea
{{\cal S}}[p, F, \hbar] =  \frac{\lambda}{\epsilon}\int_0^T dt\, d^K q \; p(t, {\bf q})  \left ( \frac{\hbar^2}{2 m}   \sum_k\frac{\partial\varphi^*}{\partial q_k} \frac{\partial \varphi}{\partial q_k}   -  i \hbar  \frac{ \partial \varphi }{\partial t}  +  {V} \right ) \label{eq:action3}
\eea
where
\bea
\varphi \equiv \log \sqrt{p} + i \epsilon \frac{F + \mu n }{\hbar}.
\eea
If $n \in \mathbb{Z}$ is indeed unobservable, then ${{\cal S}}[p, F, \hbar]$ should not depend on $n$ which is guaranteed only if
\be
\mu = \frac{2 \pi \hbar}{\epsilon}  m, \label{eq:mu}
\ee
for some $m \in \mathbb{Z}$. If that would not be true, then by studying changes in the action \eqref{eq:action3} we would be able to extract information about $n$ in a conflict with our assertion that $n$ is unobservable. In other words, ${{\cal S}}[p, F, \hbar]$ must be invariant under transformation $F  \rightarrow F + \mu n$ for all $n\in\mathbb{Z}$ which is assured if we impose the condition \eqref{eq:mu}.  To prove that $m = \pm 1$ we must now look at other discrete transformations of $F$. If $m \neq \pm 1$, then there are transformations
\be
F  \rightarrow F + \mu \frac{n}{m} \;\;\;\;\;\;\; \;\;\;\;\;\;\;\forall n \in \mathbb{Z} \label{eq:first_cond2}
\ee
which leave ${{\cal S}}[p, F, \hbar]$ invariant, but $n/m \notin \mathbb{Z}$, e.g. $n=1$, $m=2$ and $n/m=1/2$. If this is the case, then the parameter $\mu$ in \eqref{eq:first_cond} was not chosen correctly to describe the unobservability in $F$. Instead the parameter must be rescaled $\mu \rightarrow \pm \mu/m$ and then \eqref{eq:first_cond2} reduces to \eqref{eq:first_cond} and equation \eqref{eq:mu} becomes
\be
\hbar = \pm \frac{\mu \epsilon}{2 \pi }.\label{eq:second_cond}
\ee
By imposing the condition \eqref{eq:second_cond} on $\hbar$ in \eqref{eq:action3} we arrive at the Schr\"odinger action
\bea
{{\cal S}}[\Psi] =  \frac{\lambda}{\epsilon}\int_0^T dt\, d^K q \; \left ( \frac{\hbar^2}{2 m}   \sum_k\frac{\partial\Psi^*}{\partial q_k} \frac{\partial \Psi}{\partial q_k}   -  i \hbar  \Psi^* \frac{ \partial \Psi }{\partial t}  +  {V} \Psi^*  \Psi  \right ) \label{eq:action4}
\eea
where the wave function $\Psi$ is given by \eqref{eq:wavefunction}. Therefore, if the multivaluedness condition \eqref{eq:first_cond} is satisfied, then the Planck constant must be given by \eqref{eq:second_cond} and then the Schr\"odinger action \eqref{eq:action4} provides a correct statistical description of the learning dynamics of neural networks.

\section{Grand canonical ensemble}\label{sec:grand_canonical}

Consider a neural network at a learning equilibrium described by a temperature parameter, $T$, and, in addition, with a possible access to a reservoir of auxiliary neurons described by a chemical potential, $\mu$. What this means is that the learning algorithm is such that the system can either increase (i.e. neurogenesis) or decrease (i.e. neurodegeneration) the total number of active neurons, $N$. It is not immediately clear that such an algorithm would be present in an optimal learning system, but this is something that we will discuss shortly. Meanwhile, the very fact that the exact number of active neurons (or hidden variables) $N$ is unknown suggests that the system should be modeled with a grand canonical ensemble. The corresponding thermodynamic potential is the grand (or Landau) potential
\be
\Omega({\bf q}, T, \mu) =  F - \mu N
\ee
where
\be
\mu = \frac{\partial F}{\partial N}.
\ee
For a system kept at an equilibrium with constant temperature, $T$, and chemical potential, $\mu$, the fundamental thermodynamic relations is
\be
d \Omega =  dF- \mu dN  = 0.\label{eq:first_law}
\ee
The relation \eqref{eq:first_law} can be regarded as a generalization of the first law of learning that was introduced in \cite{neural_network, machine_learning} in context of a canonical ensemble of neural networks.

According to the first law \eqref{eq:first_law} the free energy, $F$, can undergo both continuous transformations due to dynamics of trainable variables, ${\bf q}$, and discontinuous transformations due to dynamics of the number of neurons, $N$. This implies that the free energy must be ``quantized'' in the following sense
\be
F({\bf q}, T, N)  =\Omega({\bf q}, T, \mu)  + \mu N. \label{eq:quant}
\ee
Since the exact number of active neurons, $N$, in the grand canonical ensemble is unknown, the free energy $F$ is only known up to an additive constant $\mu n$ where $n \in \mathbb{Z}$. If we identify  $\mu$  in \eqref{eq:first_cond} with chemical potential of the grand canonical ensemble, then the unobservability of the number of active neurons implies the multivaluedness condition \eqref{eq:first_cond}. Strictly speaking, the condition is only satisfied if the integer $n$ in  \eqref{eq:first_cond}  remains smaller than the uncertainty  in the number of  neurons
\be
 \Delta N =  \sqrt{ \langle N^2 \rangle -  \langle N \rangle^2 } = \sqrt{T\frac{\partial^2 \Omega}{\partial \mu^2} }. \label{eq:dN}
\ee
In the limit $n \ll \Delta N $ the multivaluedness condition is satisfied and the Schr\"odinger equation provides a good statistical description of the learning dynamics, but in the opposite limit $n \gtrsim \Delta N$ the Schr\"odinger description is expected to break down. However, one can argue that in an optimal neural network the parameter  $\Delta N$ must be maximized which would make the behavior of the system as quantum as possible.

Indeed, for  every ``macroscopic'' solution for trainable variables, ${\bf q}$, there is a statistical ensemble of microscopic solutions for hidden variables, ${\bf x}$. With this respect the grand canonical ensemble, $\Delta N \neq 0$, provides a clear advantage over canonical ensemble, $\Delta N =0$, as it allows for a much larger number of microscopic solutions corresponding to different values of the number of active neurons, $N$. More precisely, if the system has access to $\sim 2 \Delta N$ auxiliary neurons, then each of these neurons can either be active or not active, and then the additional entropy is given by
\be
 \Delta S \sim 2 \Delta N = 2 \sqrt{T\frac{\partial^2 \Omega}{\partial \mu^2} }.\label{eq:add_ent}
 \ee
This entropy describes the additional ``macroscopic'' solutions for trainable variables, ${\bf q}$, which can have discontinuous jumps in the free energy \eqref{eq:first_cond} due to uncertainty in the total number of neurons \eqref{eq:dN}. Therefore, an optimal neural network must be described by a grand canonical ensemble with the largest possible $\Delta N$ for which the multivaluedness condition \eqref{eq:first_cond} would be maximally satisfied. This establishes an equivalence between quantum mechanics and an optimal learning system described by a grand canonical ensemble of neural networks.

Whether the free energy $F$ (and the loss function $U = F - TS_x$) is ``quantized'' \eqref{eq:quant} and whether it can change discontinuously depends on the learning system. In Ref. \cite{machine_learning} it was shown numerically that for the bulk loss function the discontinuous jumps are suppressed, but for the boundary loss function the discontinuous jumps are expected even when the total number of neurons is kept constant. This result agrees very well with our analysis of the grand canonical ensembles and with the first law of learning \eqref{eq:first_law}. Indeed, from the point of view of the bulk neurons the total number of neurons is constant, the relevant statistical ensemble is canonical and the discontinuous jumps do not occur,
\be
dU = T dS_x.
\ee
 On the other hand, from the point of view of only boundary neurons, some of the bulk neurons can act as a reservoir of auxiliary neurons, the relevant ensemble is grand canonical and the discontinuous jumps are expected,
 \be
 dU = T dS_x + \mu dN.
 \ee
Therefore, in addition to theoretical considerations we also have preliminary numerical results suggesting that the discontinuous jumps in the boundary loss function is a result of the learning dynamics described by a grand canonical ensemble.

Using the ``quantization'' of the free energy \eqref{eq:quant} and the optimal value for the Planck constant \eqref{eq:second_cond}, the wave function \eqref{eq:wavefunction} can written as
\be
\Psi\left ( t, {\bf q} \right ) =  \sqrt{p\left ( t, {\bf q} \right )} \exp \left ( \frac{i \Omega\left ( t, {\bf q} \right ) \epsilon}{\hbar} \right ). \label{eq:wavefunction2}
\ee
As an example, consider a grand potential which can be expressed as a sum of a fixed time-independent term and a time-dependent term, i.e.
\be
\Omega\left ( t, {\bf q} \right )  = \Omega_0\left ( {\bf q} \right ) +\Omega_1\left ( t, {\bf q} \right ).
\ee
In such limit the Sch\"rodinger action \eqref{eq:Schrodinger} can be rewritten as
\bea
{{\cal S}}[\tilde{\Psi}] =  \frac{\lambda}{\epsilon}\int_0^T dt\, d^K q \; \left (- \frac{\hbar^2}{2 m}\tilde{\Psi}^*  \sum_k  \left ( \frac{\partial}{\partial q_k} + i \frac{e A_k}{\hbar} \right )  \left ( \frac{\partial }{\partial q_k}  + i \frac{e A_k}{\hbar}  \right ) \tilde{\Psi}  -  i \hbar  \tilde{\Psi}^* \frac{ \partial \tilde{\Psi} }{\partial t}  +  {V} \tilde{\Psi}^*  \tilde{\Psi} \right ) \notag\label{eq:action5}
\eea
where the new wave function is now defined as
\be
\tilde{\Psi}\left ( t, {\bf q} \right ) \equiv \sqrt{p\left ( t, {\bf q} \right )} \exp \left ( \frac{i \Omega_1\left ( t, {\bf q} \right ) \epsilon}{\hbar} \right ) \label{eq:wavefunction3}
\ee
and
\be
A_k\left ( {\bf q} \right ) \equiv \frac{\epsilon}{e} \frac{\partial \Omega_0\left ( {\bf q} \right )}{\partial q_k}.
\ee
Then upon variation of the action with respect to the new wave function we get
\be
-  i \hbar \frac{\partial }{\partial t } \tilde{\Psi}  =  \left [ \frac{\hbar^2}{2 m} \left ( \frac{\partial}{\partial q_k} + i \frac{e A_k}{\hbar} \right ) \left ( \frac{\partial }{\partial q_k}  + i \frac{e A_k}{\hbar}  \right )      -  V \right ] \tilde{\Psi}. \label{eq:Schrodinger2}
\ee
Note that the result is only valid when the time-independent term, $\Omega_0\left ( {\bf q} \right )$, is fixed (i.e. is not varied), which is exactly the limit in which the Schr\"odinger equation \eqref{eq:Schrodinger2} provides a good description of a quantum particle with ``charge'' $e$ in an external field described by the ``vector potential '', ${\bf A}$.

\section{Quantum superpositions}\label{sec:superposition}

In the previous sections we modeled the learning dynamics of neural networks using either Madelung \eqref{eq:neural_FP}, \eqref{eq:neural_NS} or Schr\"odinger \eqref{eq:Schrodinger} equations. In both cases the dynamics was described in a position basis ${\bf q}$ which is the preferred basis in our entire construction. The main difference between Madelung and Schr\"odinger equations is that the Schr\"odinger dynamics is linear and thus can be expressed in any orthonormal basis without loosing the generality. What is less clear is how to put the system in an arbitrary initial state or how to measure the system with respect to arbitrary measurement operators. In particular, it is not clear how to start the dynamics in a superposition of position eigenstates or how to perform measurements using non-diagonal (in the position basis) measurement operators.

The concept of measurement plays a central role in quantum physics, as especially emphasized in Bohr's complementarity principle \cite{bohr,wheeler}. What we deal with is never a ``quantum system by itself'' but a result of its interaction with some measurement devices, and we can choose different descriptions choosing different sets of devices. For example, we can measure either coordinate of a particle by a local detector which clicks when the particle interacts with it or momentum of the particle via its wave properties, using to this aim diffraction lattices, etc. For the case of neural networks the coordinate representation is special. It is very straightforward to measure the set of ${\bf q}$ at a given time instant, this information is directly available at the computer. At the derivation of Schr\"odinger equation in logical inference approach \cite{logical_inference1} the space is supposed to be filled by coordinate detectors as well whereas measurement of momenta is not so easily realizable even as a gedanken experiment. Note however that the quantum mechanics allows a purely space-time formulation which was realized in Feynman's path integral approach \cite{feynman_hibbs}. Mathematically speaking, the solution of the Cauchy problem for the Schr\"odinger equation can be presented as a path integral, via splitting of the evolutionary operator into many elementary factors with the further use of Trotter decomposition formula \cite{schulman}. All interference phenomena, energy quantization and other specifically quantum phenomena follow immediately from this representation \cite{feynman_hibbs,schulman,kleinert}. Importantly, the trajectories giving the main contribution to the path integral are continuous but not continuously differentiable \cite{berezin} which means impossibility of simultaneous measurements of coordinates and velocities. Coming back to the operator language, one can discuss the measurements of noncommutative coordinate operators at different time instants rather than the measurements of noncommutative coordinate and velocity operators at the same time instant. Therefore furthe we will discuss only measurements of the coordinates ${\bf q}$.   

In artificial neural networks numerical values of the trainable variables ${\bf q}$ are only known up to numerical precision and so after measurement in a position basis the system can only be in one of a finite number of states, i.e. ${\bf q} \in \{{\bf q}_1, {\bf q}_2, ...,{\bf q}_M\}$. Using the bra-ket notations the position eigenstates are given by
\be
|{\bf q}\rangle \in \{|{\bf q}_1\rangle, |{\bf q}_2\rangle, ...,|{\bf q}_M\rangle\},
\ee
and the most general initial state can be expressed as a linear superposition 
\be
|\Psi(0)\rangle = \sum_{i=1}^M \Psi(0,{\bf q}_i)  |{\bf q}_i\rangle.\label{eq:pure}
\ee
It is certainly possible to use a random number generator to set the initial state to be in a state $|{\bf q}_i\rangle$ with probability $p_i = |\Psi(0,{\bf q}_i)|^2$, but then the system would not be in a pure state \eqref{eq:pure}. Such states are known as mixed states that are usually described by a density matrix
\be
\hat{\rho} = \sum_{i=1}^M p_i |{\bf q}_i\rangle \langle {\bf q}_i|. \label{eq:mixed}
\ee
It seems that the only way for a system to remain in a pure state is through unitary evolution which can in general be time-dependent. Then to prepare a superposition state  \eqref{eq:pure} at time $t=0$ we can pre-evolve it starting from a position eigenstate $|{\bf q}_j\rangle$ at time $t=-t_-$, i.e.
\be
|\Psi(0)\rangle = e^{-i t_-\hat{H}_-/\hbar} |{\bf q}_j\rangle
\ee
where $\hat{H}_-$ is the pre-evolution Hamiltonian operator. Note that $\hat{H}_-$ emerges from a microscopic loss function and a training dataset which need not be the  same as for $\hat{H}$ which governs the main part of the evolution
\be
|\Psi(T)\rangle = e^{- i T \hat{H}/\hbar} |\Psi(0)\rangle. 
\ee
It is important to emphasize that although the loss functions, training datasets and emergent Hamiltonians for pre-evolution and main evolution can differ, the neural architectures, described by trainable ${\bf q}$ and non-trainable ${\bf x}$ variables, must remain the same. Of course, this does not guarantee that we can use the pre-evolution to prepare all possible superposition states, but by modifying $t_-$ (or pre-evolution time interval) and  $\hat{H}_-$ (or pre-evolution loss function and training dataset) a larger variety of pure initial states can be realized. Also note that on top of realizing superposition states one can still use a random number generator to create a mixed state by starting the pre-evolution with \eqref{eq:mixed} and then the density matrix at time $t=0$ would be
\be
\hat{\rho}(0) = e^{-i t_- \hat{H}_-/\hbar} \left ( \sum_{i=1}^M p_i    |{\bf q}_i\rangle \langle {\bf q}_i| \right ) e^{i t_- \hat{H}_-/\hbar} 
\ee
which need not be diagonal.

What about measurement operators? Can we use a similar method to (effectively) measure the system using non-diagonal measurement operators, 
\be
\hat{O}_{m} \equiv \sum_{i,j=1}^M O^{(m)}_{ij} |{\bf q}_i\rangle \langle {\bf q}_j|?
\ee
If the quantum description is correct then the probability of observing a given measurement operator  must be given by
\be
p(m) = \langle \Psi(T)|  \hat{O}_{m}^\dagger \hat{O}_{m}   |\Psi(T)\rangle
\ee
where 
\be
\sum_m \hat{O}_{m}^\dagger \hat{O}_m  = \hat{I}. 
\ee
For diagonal measurement operators (denoted by $\hat{D}_{m}$'s) the probabilities are given by
\be
p(m) =\sum_{i=1}^M   \left | D^{(m)}_{ii}\right |^2 \left | \Psi(T,{\bf q}_i)\right |^2 \label{eq:diag}
\ee
and by measuring positions ${\bf q}_i$'s, probabilities $ |\Psi(T,{\bf q}_i)|^2$'s can be calculated and $p(m)$'s can be verified against theoretical predictions. However, for non-diagonal measurement operators
\be
p(m) =\sum_{i=1}^M   \left | \sum_{j=1}^M O^{(m)}_{ij} \Psi(T,{\bf q}_j)  \right |^2,
\ee
and the knowledge of probabilities  $ |\Psi(T,{\bf q}_i)|^2$ is not sufficient for calculating $p(m)$'s or for performing measurements of the corresponding operators. On the other hand, what one can do is to post-evolve the state $|\Psi(T)\rangle$ to $|\Psi(T+t_+)\rangle$ using some post-evolution Hamiltonian $\hat{H}_+$, i.e.
\be
|\Psi(T+t_+)\rangle = e^{- i t_+ \hat{H}_+/\hbar} |\Psi(T)\rangle,
\ee
and then measure it using some set of diagonal operators $\hat{D}_{m}$'s. Then the probabilities of measuring $\hat{D}_{m}$'s would be given by \eqref{eq:diag}, which can be calculated and verified against theoretical predictions, but the same probabilities can also be expressed as
\bea
p(m) &=& \langle \Psi(T+t_+)|  \hat{D}_{m}^\dagger \hat{D}_{m}   |\Psi(T+t_+)\rangle\\
 &=&  \langle \Psi(T)| e^{i t_+ \hat{H}_+/\hbar} \hat{D}_{m}^\dagger e^{-i t_+ \hat{H}_+/\hbar}e^{i t_+ \hat{H}_+/\hbar} \hat{D}_{m}   e^{- i t_+ \hat{H}_+/\hbar} |\Psi(T)\rangle\\
 &=&  \langle \Psi(T)|  \hat{O}_{m}^\dagger \hat{O}_{m}  |\Psi(T)\rangle,
\eea
i.e. as probabilities of effectively  measuring non-diagonal operators 
\be
\hat{O}_{m}  = e^{i t_+ \hat{H}_+/\hbar} \hat{D}_{m}   e^{- i t_+ \hat{H}_+/\hbar}.
\ee
Of course, there is no guarantee that we would be able to effectively measure all possible sets of the measurement operators since the procedure is still limited by possible choices of the post-evolutionary Hamiltonian $\hat{H}_+$ (or loss function and training dataset) and time interval $t_+$.

\section{Discussion}\label{sec:discussion}

In this paper we analyzed the emergent macroscopic dynamics of neural networks, but deliberately omitted specifications of the microscopic dynamics. The only important microscopic ingredient was that there are two types of degrees of freedom which correspond respectively to trainable and hidden variables. In the emergent picture the trainable variables were identified with the variables of the wave function, but the hidden variables were only described at the level of statistical ensembles. In fact, it was not even important whether the hidden variables are actually non-trainable or only appear to evolve as non-trainable variables, but what was important is that their statistics can be described with a grand canonical ensemble. In this respect one can argue that the emergent quantumness is a generic macroscopic prediction of any learning system with a coupled dynamics of these two types of variables, i.e. trainable and hidden.

It is also worth emphasizing that the trainable variables were assumed to be continuous and the configuration space was assumed to be flat. More generally some of the variables (hidden or trainable) can be either discrete or the configuration space can be curved. In such cases the derivatives would be replaced with either finite differences or with covariant derivatives, but this would not alter the main conclusion of the paper. In fact, we expect that in more realistic models of neural network (which could give rise to emergent quantum field theories and gravity) the trainable variables must include features of both discrete variables and curved spaces. On the other hand, the {\it discreteness} of the neural network (i.e. the number of neurons is a discrete integer) is a crucial point of the whole construction and for the main result, i.e. the emergence of quantumness.  This result has some immediate and important implications for machine learning, physics and biology that we shall discuss next.

{\it Machine learning.} Machine learning is perhaps the simplest example of a learning system which has some apparent advantages over physical and biological systems. Artificial neural networks are well defined mathematically and, as such, provides an excellent experimental platform for testing the new ideas numerically. There are at least three (not unrelated) ideas which follow directly from our results. According to our analysis an optimal learning system should be based on an algorithm which allows for the number of hidden variables to vary. One way to allow the number of hidden variables to change is to develop an algorithm designed specifically for addition and removal of the auxiliary neurons. In this respect it might be useful to develop a {\it neurodegeneration} method for removing neurons that causes the smallest increase in the loss function and a {\it neurogenesis} method for adding auxiliary neurons that causes the largest decrease in the loss function.

Another possibility is to develop an algorithm in which the change in the number of hidden variables would occur dynamically. We expect that this is what might actually be happening behind the scenes in deep learning. If correct, this suggests a simple and intuitive explanation of why the deep neural networks (with many hidden layers) are very effective in learning. The reason might be that the bulk neurons on the hidden layers can act as a reservoir  of auxiliary neurons and the corresponding ensemble becomes grand canonical. In such a limit the correct effective description of the learning dynamics of the boundary system is quantum with all of the computational advantages which come with it.  Perhaps a lot more importantly, the emergent quantumness implies that one might be able to design an artificial neural network which can mimic the behavior of a quantum computer. Of course, such an artificial quantum computer would not be quantum per se, but one could still make it maximally quantum by designing an algorithm which maximizes the uncertainty in the number of active neurons.

{\it Physics.} We can now try to tackle the somewhat more difficult problem of modeling physical systems using neural networks. Indeed, if quantum mechanics provides a good description of the physical world and a good description of the neural networks, then why cannot the physical word be a neural network? This was precisely the questions that was asked in Ref. \cite{neural_network} where not only quantum mechanics, but also gravity and observers were described as emergent phenomena. (See Refs. \cite{Jacobson,Verlinde, Padmanabhan,DSV} for other approaches to emergent gravity). In this paper we concentrated mostly on the emergent quantum behavior of the neural networks, but our results have some interesting implications for both gravitational and biological systems. Our main quantum result is that the correct statistical ensemble of hidden variables is the grand canonical ensemble, where the chemical potential is what determines the valued of the physical ``Planck constant''. In the quantum limit the learning system satisfies the following two conditions:

 (1) the system is at a learning equilibrium (i.e. small gradient of the free energy) and

 (2) the quantum phase is multivalued  (i.e. large uncertainty in the number of neurons).  \\
However, since the emergent quantum behavior is only approximate, it is also important to identify the systems in which significant deviations from the quantum behavior are expected. For example, for a canonical ensemble of hidden variables the conditions (1) can be satisfied, but the condition (2) is badly violated and then the system is better described with the Madelung equations. In an opposite limit, when the condition (2) is satisfied, but the condition (1) is violated, the system is better described with the Hamilton-Jacobi equations (see Ref. \cite{neural_network} for details).

Perhaps a more surprising aspect of the learning dynamics is that one might be able to derive a couple of dual descriptions of the very same system. In a ``boundary'' description one keeps track of only a small number of trainable variables which have already thermalized, i.e. satisfying the condition (1), and the rest of the variables are treated as hidden variables whose total number is unknown, i.e. satisfying the condition (2). In a ``bulk'' description one keeps track of most of the trainable variables which can be very far from the true equilibrium,  i.e. violating the condition (1),  and the total number of hidden variables is small and its fluctuations are also small, i.e. violating the condition (2). For the boundary system the correct emergent description is quantum, but for the bulk system the correct description is mostly classical and in some cases gravitational \cite{neural_network}. This resonates well with a holographic conjecture which states that a gravitational system in the bulk should have a quantum dual description on the boundary \cite{Witten,Susskind,Maldacena}. Indeed, if the microscopic neural network is being trained by processing the new training data through its boundary (as for example in the deep feedforward neural networks), then the boundary neurons should be the first to thermalize. Moreover, from the point of view of the boundary neurons the ensemble of hidden variables is in a grand canonical equilibrium, and then the emergent dynamics is quantum. On the other hand, the bulk neurons would be further away from the equilibrium and their emergent dynamics would be mostly classical and, perhaps, in some limits gravitational \cite{neural_network}. This offers an interesting new perspective on the holographic principle and on gravity as an emergent phenomenon. (See also Ref. \cite{Dvali} for an alternative approach to gravity and holography in context of neural networks.)

The emergent quantumness also provides a new twist to the long-standing problem of derivation of irreversible macroscopic laws from reversible microscopic ones. The basic equations determining the dynamics of neural network  are already irreversible, due to the presence of diffusion terms with real diffusion coefficient (contrary to imaginary diffusion coefficient in time-reversal symmetric Schr\"odinger equation), and it is the reversibility of the microscopic laws turns out to be an emergent phenomenon, due to negative entropy production during learning. The negative entropy production  is the direct consequence of the second law of learning, i.e. the total entropy can never increase during learning and is constant in the learning equilibrium \cite{machine_learning, neural_network}. Then, after the equilibrium is reached, any positive entropy production due to diffusion of trainable variables must be  balanced by the negative entropy production of either trainable or hidden variables.  On the shortest time scales the interplay between positive and negative entropy productions can be expressed explicitly by including respectively the diffusion and drift terms in the Fokker-Planck equation, but on the longer time scales the entropy balance is expressed as an optimization problem which can be described by the Schr\"odinger equation. Therefore, for the long time scale dynamics there is an emergent time reversal symmetry, but on the short time scales the symmetry must be broken.

{\it Biology.}  From a biological perspective, it is very tempting to speculate whether our mind can be modeled as a neural network which undergoes a learning evolution. In that case, its behavior near equilibrium should indeed resemble the quantum systems since it would be described by an effective Schr\"odinger equation. Of course, the effective Planck constant arising in such an equation has nothing in common with the real physical Planck constant. In this sense, our approach is essentially different from those of Refs.\cite{Penrose,Fisher} where the relevance of truly quantum processes in our bodies for our mind is suggested. The other point worth to be emphasized is that, in our approach, only a fully optimized neural network has this property of ``quantumness''. Of course, it is not obvious at all whether our real brains originated from a real biological evolution on a specific planet are optimal enough to be ``quantum''. Any suggestions of such kind would be unavoidably very speculative, but, in our opinion, deserve to be considered.

Speaking more generally on the biological evolution, one should mention Ref.\cite{Koonin} where the question of what kind of physics would be needed to describe evolution is discussed. The key point is a coexistence of two levels (genotype and phenotype) which is essentially entangled in a very unusual way from the point of view of conventional statistical mechanics. Physical carriers of genetic information are (very roughly speaking) macromolecules subjected to thermal fluctuations, electrostatic interactions, with electronic structure determined by quantum mechanics, etc. However, the functionality of the genetic information carriers cannot be adequately described in terms of their physical properties only. They are projected to a phenotype level, and at this level are subjected to selection with the laws which are also in agreement with general laws of physics and chemistry (as we believed), but act on a completely different level of macroobjects. This was compared with the role played by von Neumann measurements due to interaction with macroscopic measuring devices in quantum mechanics \cite{Koonin}. In this regard, the concept of emergent quantumness via neural network approach developed here might be useful for specification and formalization of this, still vague and preliminary, analogy. Anyway, there seems to be a clear association of genotype-phenotype duality in biology to the duality of hidden and trainable variables in neural networks. Another thought-provoking question is the utility of ``quantum jumps'' in the fitness values in the ``quantum-like'' evolutionary dynamics, but it obviously goes far beyond the scope of this particular work and deserves a separate consideration.

{\it Acknowledgments.}  VV was supported in part by the Foundational Questions Institute (FQXi). MIK acknowledges a support by NWO via Spinoza Prize.

\end{document}